\documentclass[12pt]{article}

\def\lromn#1{\uppercase\expandafter{\romannumeral#1}}

\begin{document}

\begin{center}
\begin{large}
\textbf{
Photon Irradiated Enhancement
as a Tool of Investigating Fundamental
Physics beyond Standard Model
}

\end{large}
\end{center}

\vspace{2cm}
\begin{center}
\begin{large}
M. Yoshimura

Department of Physics, Okayama University \\
Tsushima-naka 3-1-1 Okayama
Japan 700-8530
\end{large}
\end{center}

\vspace{4cm}


\begin{center}
\begin{Large}
{\bf ABSTRACT}
\end{Large}
\end{center}

We clarify how intense laser irradiation
leads to an enhancement of rare processes that may occur within atoms.
Non-perturbative calculation using a coherent laser beam
gives an exact, time dependent formula of the
enhancement factor in the large power limit.
A high quality laser may provide a new tool of 
experimentally investigating physics far beyond the standard model
of particle physics such as lepton and baryon nonconservation.

\newpage

In our recent paper \cite{insty} it was pointed out that intense laser beam,
when irradiated to appropriate heavy atoms, 
can enhance rare processes related
to atomic electron capture by nucleus, otherwise difficult to detect.
In the present work we extend the formalism such that
non-perturbative effects of laser-atom interaction
are fully taken into account
by directly solving dynamics of laser irradiation onto target atoms.
In the dipole approximation of the two level problem of atoms
the result suggests an onset of a repeated process 
of compression and expansion of the atomic electron cloud.
Rare processes that may subsequently
occur via the overlap of atomic electrons with 
nucleus may thus be enhanced.
We propose to call this compression mechanism as PHIRAC
to abbreviate PHoton IRrAdiated Compression of the electron cloud.
For magnetic transitions such as those between hyperfine split levels this
interpretation of electron cloud compression is not appropriate,
but one can obtain a large enhancement of rare processes as well.

Our discussion in the present work is independent of 
any particular rare process $X$ subsequent to photon absorption.
It may thus provide a new method
of exploring physics far beyond the standard model of particle physics.
In this way one can experimentally explore 
lepton number nonconservation (LENNON) of the type $e^- \rightarrow e^+$, 
baryon number nonconserving (BARNNON) process of the kind
$e^- + N \rightarrow$ many $\pi'$s, 
and hopefully many other fundamental processes
that face physics beyond the standard model.

Consider laser irradiation which induces transitions 
between two electronic levels of a target atom,
denoted by $| e \rangle $ and $| g \rangle$.
We imagine that the target is irradiated continuously.
For instance putting the target into a resonator may be useful.
The system of laser and atom
is then described by a Hamiltonian of the form
\cite{laser book},
\begin{equation}
H = \frac{\omega_{eg}}{2}\sigma_3 + \omega a^{\dagger}a
+ \tilde{s}( \sigma_+ a  +  \sigma_- a^{\dagger})
\,.
\end{equation}
Here the Pauli matrices 
$\sigma_i$ act on two levels of the ground $| g \rangle$ and the excited 
$| e \rangle$ state, and $\omega_{eg} = \omega_e - \omega_g$ is
the energy level difference.
We made what is called the rotating wave approximation
in the literature \cite{laser book},
by neglecting $\sigma_+ a^{\dagger}$ and its conjugate,
which should be justified for mostly energy-conserving processes.
The coupling strength $s$ is given by 
\begin{equation}
\langle e | \tilde{s} \sigma_+ | g \rangle \equiv
s = - \langle e | \vec{d} | g \rangle \cdot \vec{e} 
\,, \hspace{0.5cm}
\vec{e} = i\frac{\omega}{\sqrt{2\omega V}}\vec{\epsilon}_k
\end{equation}
for the dipole transtion, with
$\vec{d}$ the dipole operator of atomic electron and $\vec{e}$
the electric field of a single photon beam with polarization 
$\vec{\epsilon}_k$.
When multiplied by a photon number $N$, this strength
is expressed as
\( \:
Ns^2 = \pi \gamma_d P/\omega^3
\: \)
where $P$ is the laser power in the unit of energy/
(time $times$ area). 

The Hamiltonian is block-diagonal, hence
the effect of laser irradiation
can be solved by decomposing the infinite dimensional Fock
space and using a mixture of two states, 
$| e\,, n \rangle = (a^{\dagger})^{n}/\sqrt{n!}|e \rangle$ 
and $| g\,, n \rangle = (a^{\dagger})^{n}/\sqrt{n!}|g \rangle$;
\( \:
| \psi (t) \rangle = \sum_n (c_{g\,n+1}(t) | g\,, n +1 \rangle 
+ c_{e\,n}(t)| e\,, n \rangle)
\,.
\: \)
The Schroedinger equation to be solved is
\begin{eqnarray}
&&
\hspace*{-1cm}
i \frac{d}{dt} \left(
\begin{array}{c}
  c_{e\,n} \\
  c_{g \,n+1 }
\end{array}
\right)
= \cal{H} 
\left(
\begin{array}{c}
  c_{e\,n} \\
  c_{g \,n+1 }
\end{array}
\right)
\,, \hspace{0.5cm}
\cal{H}
= \left(
\begin{array}{cc}
 n\omega + \frac{\omega_{eg}}{2} &  s \sqrt{n+ 1}\\
 s \sqrt{n+ 1} & (n + 1) \omega - \frac{\omega_{eg}}{2}
\end{array}
\right)
\,.
\end{eqnarray}
The solution may be written in terms of what is called dressed states
denoted by $| \pm \,, n\rangle$;
assuming a spacially constant (valid in the long wavelength approximation) 
laser field, the Hamiltonian diagonalization is possible with
\begin{eqnarray}
&&
| +\,, n\rangle =
\cos \frac{\varphi_n}{2} | e\,, n \rangle + \sin 
\frac{\varphi_n}{2} | g\,, n+1 \rangle
\,,
\\ &&
| -\,, n\rangle =
- \sin \frac{\varphi_n}{2}  | e\,, n \rangle +  \cos \frac{\varphi_n}{2}
| g\,, n+1 \rangle
\,,
\end{eqnarray}
where
\begin{eqnarray}
&&
\tan \varphi_n = \frac{2s \sqrt{n+1}}{\omega - \omega_{eg}}
\,,
\hspace{1cm}
\omega_{\pm} =
(n + \frac{1}{2})\omega \pm \frac{\Omega_n}{2}
\,, 
\end{eqnarray}
with $\Omega_n = \sqrt{(\omega - \omega_{eg})^2 + 4s^2 (n+1) }$ the Rabi
frequency \cite{laser book}.
Unless the photon energy $\omega$ is very far from the
resonance energy $\omega_{eg}$, the mixing is nearly maximal;
\( \:
\sin^2 \varphi_n \approx 1
\,.
\: \)
We thus assume the maximal mixing for simplicity; $\varphi_n = \pi/2$.

It is reasonable under the continuous laser irradiation to assume 
that the target is initially in a superposed state of two levels,
\(
\:
| \frac{1}{2} \rangle \equiv 
\frac{1}{\sqrt{2}} (e^{i\delta}| g \rangle +  | e \rangle)
\:
\)
with $\delta = \pi/2$.
The result is insensitive to the choice of this phase and
the initial condition as a whole.

Time evolution is then given by 
\begin{eqnarray}
&&
\langle g\,, n+ 1| \frac{1}{2}\,, n\, ; t \rangle
= i e^{-i (n + 1/2)\omega t}
\cos(\frac{\Omega_n t}{2} + \frac{\pi}{4})\, c_{n}^{(\gamma)}(0)
\,.
\end{eqnarray}
In the rest of this paper we take the coherent state of laser, thus
\( \:
c_{ n}^{(\gamma)}(0) = e^{- N/2}N^n/\sqrt{n!}
\,,
\: \)
where the average photon number $\langle n \rangle = N$
and the dispersion $\langle (\Delta n)^2 \rangle = N$.

It is conceptually important to
distinguish from the Schroedinger (S) picture and use the interaction (I)
picture, since the subsequent rare process is treated in the
perturbation theory.
Fortunately, since the laser-atom system is exactly solvable,
one may straightforwardly use the identity 
\( \:
_{S}\langle a | b \rangle_{S} = _{I}\langle a | b \rangle_{I}
\,,
\: \)
to simplify computation.

We imagine a circumstance under which rare processes occur
via the ground state $|g \rangle$ of zero angular momentum, a $ms$ state.
It is assumed that at a time $t$
the electron in the excited $|e \rangle$ 
goes to the $ms$ $|g \rangle$ state due to the stimulated emmision, and
is subsequently captured by
nucleus, with a probability proportional to the wave function
factor $|\psi_{ms}(0)|^2$.
An example of subsequent rare processes of this sort is
LENNON electron capture of the kind $e^- \rightarrow e^+$, as discussed
in \cite{insty}.
The probability amplitude of laser irradiated rare process is then given by
\begin{eqnarray}
&&
\hspace*{-1cm}
\tilde{\cal M}_{1/2}(t) = e^{- (\gamma_e + \gamma_g)t/4}
\sum_{n=0}^{\infty}\,
{\cal M}_X(t)
\frac{s \sqrt{n+ 1}}{\omega - \omega_{eg} + i \gamma/2}
\langle g\,, n+ 1| \frac{1}{2}\,, n\, ; t \rangle
\label{probability amplitude of phirac}
\end{eqnarray}
where ${\cal M}_X(t)$ is the amplitude for the rare $X$ process
from the $ms$ state.

Computation of the discrete $n$ sum (\ref{probability amplitude of phirac}) 
is well approximated in the large $N$ limit
by a continuous $n$ integral.
Using the large $n$ limit 
formula of $n!$, and the coherent state expression for $c_{ n}^{(\gamma)}(0)$,
the integrand is found to change violently, and one
may estimate the integral by a gaussian approximation 
around the stationary phase, or the saddle point.
The saddle point $n_0$ of the integrand is determined by
the minimal variation
of the exponent of the integrand and is given by taking the 
$n-$derivative of the exponent to vanish. A complex saddle is thus
obtained;
\(
\:
n_0 \approx N e^{-2i \omega t}
\,.
\:
\)
Making the gaussian approximation around this saddle gives
the rate formula,
\begin{eqnarray}
&&
\hspace*{1cm}
|\tilde{{\cal M}}_{1/2}(t)|^2 \approx 
(\frac{\pi}{2})^{1/2}N^{3/2} 
\frac{s^2\, |{\cal M}_X(t)|^2}{(\omega - \omega_{eg})^2 + \gamma^2/4} 
\nonumber \\ &&
\hspace*{-1cm}
\exp [- N (1 - \cos 2\omega t) - (\gamma_e + \gamma_g)t/2]
[\sinh^2 (\sqrt{N} st \sin \omega t ) + \cos^2 (\sqrt{N} st \cos \omega t)]
\,.
\label{time evolving rate}
\nonumber \\ &&
\end{eqnarray}

As a funtion of time, the rate becomes very large of order 
\( \:
N^{3/2}
\,,
\: \)
periodically at
$\cos 2\omega t = 1$.
Outside regions of a time range $\approx 1/(\omega \sqrt{N})$ 
the rate is very small, of order 
$N^{3/2} e^{- N} $.
Thus, the rate is sizable only 
around infinitely many discrete times of $t = k\pi /\omega $,
with $k$ any positive integer.
In other words, the rate has a spiky time profile with a period
$\pi/\omega$.
A formula valid for $N \gg 1$
\begin{eqnarray}
&&
\hspace*{-1.5cm}
\exp[- N(1 - \cos 2\omega t )]
\approx  \sqrt{\frac{\pi}{2}}\frac{1}{\omega \sqrt{N}}
\sum_k
\delta (t - \frac{k \pi }{\omega})
\,,
\end{eqnarray}
may then be used.
For $\omega t \gg 1$, it is reasonable to take a time average over
$\Delta t \approx \pi/\omega \times$ a few, which gives 
a time variant averaged rate,
\begin{eqnarray}
&&
\tilde{{\cal R}}_{1/2}(t) = 
\frac{N}{2}\, \frac{s^2\,e^{- (\gamma_e + \gamma_g)t/2}}
{(\omega - \omega_{eg})^2 + \gamma^2/4}
\,{\cal R}_X(t)
\,,
\label{rate formula}
\end{eqnarray}
with
\( \:
{\cal R}_X(t) = d|{\cal M}_X(t)|^2/dt
\,.
\: \)

The last factor $ {\cal R}_X(t)$ 
may differ in rare processes in which one is interested.
For exmaple, the LENNON conversion of the type $e^- \rightarrow e^+$  has
\( \:
{\cal R}_{e^- \rightarrow e^+}(t) =
|\psi_{ms}(0)|^2 \sigma_{e^- \rightarrow e^+}
\,,
\: \)
where $\sigma_{e^- \rightarrow e^+}$ is the cross section of
free electron capture, a virtual process considered for our
gedanken experiment.
This quantity is computed using the perturbation theory.

One may summarize arguments so far by
defining a quality factor $Q(\omega)$ which signifies
the rate enhancement,
\begin{eqnarray}
&&
Q \equiv r\,\frac{\pi}{2} \frac{\gamma_d}{\omega_0^2 }
\int d\omega \frac{ {\cal P}(\omega)}
{ \omega [(\omega - \omega_0)^2 + \gamma^2/4]}
\,,
\label{q-general}
\end{eqnarray}
where
\(\:
\int d\omega {\cal P}(\omega) 
= P
\:\)
is the total laser power in the unit of energy /(time $\times$ area).
The factor $r $ is the wave function ratio sqaured, for instance,
for LENNON 
\begin{eqnarray}
&&
r = \frac{|\psi_{ms}(0)|^2}{|\psi_{ns}(0)|^2}
= (\frac{r_{ns}}{r_{ms}})^3
\approx O[(\frac{n}{m})^6]
\,, \hspace{1cm}
n = 1
\,.
\end{eqnarray}
For a laser beam of the energy resolution $\Delta E \gg \gamma$
we may replace 
\(\:
1/[(\omega - \omega_0)^2 + \gamma^2/4]
\:\)
by
\(\:
\frac{2\pi }{\gamma} \delta (\omega - \omega_0)
\:\)
for a laser beam of Lorentzian energy distribution.
When the laser tuning is perfect, 
\(\:
{\cal P}(\omega_0) \approx P/\Delta E
\,.
\:\)
It is thus found that
\begin{eqnarray}
&&
Q \approx r\,
\frac{\pi^2  P}{ \omega_0^4 }\frac{\omega_0}{\Delta E}
\approx
1.6 \times 10^{6}\, r\,\frac{P}{W\, mm^2}
(\frac{\omega_0}{eV})^{-4}(10^{-9}\frac{\omega_0}{\Delta E})
\,.
\label{quality factor}
\end{eqnarray}
Note a strong dependence on the photon
energy $\propto \omega_0^{-4}$, which should be important
to get a large quality factor for BARRNON.
This formula for the quality (\ref{quality factor}) agrees with the result of
\cite{insty} after a minor correction \cite{4-pi}.
The spiky time profile is however a result of the present nonperturbative
formalism.

In order to advance an intuitive understanding of the enhancement
mechanism, it might be instructive to compute the electron displacement
squared
\begin{eqnarray}
&&
(\delta {\cal D})^2 (t)
= d^2 \, \sum_n 
\langle \frac{1}{2}\,, n \,; t
|\sigma_- \sigma_+ a^{\dagger} a |\frac{1}{2}\,, n \,; t \rangle
\nonumber
\\ &&
\hspace*{1cm}
\approx
\frac{d^2}{2} \, \sum_n (1 - \sin \Omega_n t) |c_n^{(\gamma)}(0)|^2
\,.
\end{eqnarray}
The leading term of $O[d^2 N/2]$
simply shows that quantum mechanics gives a result consistent
with the classical Lorentz oscillator model.

The behavior of the next leading term in the large $N$ limit 
of $(\delta {\cal D})^2 (t)$ and
the linear dipole ${\cal D}(t)$ is more complicated.
They exhibit a spiky time profile in much
the same way as the rate formula, but with an important difference
of time scale;
these quantities that have classical analogues vary in time much more
slowly, the spike interval being given by $4\pi / \Omega_R$ with
$\Omega_R = 2 s \sqrt{N}$ the Rabi resonance frequency.
We may only say that the spiky time profile observed in the
rate enhancement is a signal of the onset of the oscillatory
behavior of the electron displacement.
The link is however indirect.

Baryon nonconservation may experimentally be investigated by searching for
the atomic electron capture of the type \cite{nakano},
\( \:
e^- + N \rightarrow \pi + \pi
\,.
\: \)
Enhancement factor $\propto P$ may compensate the small 
nuclear overlap factor of order $(a_B m_{\pi})^3 \approx 10^{-15}$
of atomic electrons that otherwise disfavors this process.
A rough estimate of the rate gives the enhancement factor of order,
\(\:
Q\,(r_A A^{1/3} m_{\pi})^{-3}
\,,
\: \)
where the pion mass $m_{\pi}$ times $A^{1/3}$
gives a measure of the inverse nuclear size,
and $r_A$ is the atomic size.
It is important to go to a low frequency range for a large
enhancement of order $10^{25}$.
Thus, the use of the Zeeman split hyperfine levels is promissing,
which involves the microwave region.

In summary, the prospect of search for new physics far beyond the standard
model appears bright if one uses an appropriate high quality laser.

\vspace{1cm}
I would like to thank Y. Kuno, I. Nakano, and N. Sasao for
stimulating discussions.


\begin{thebibliography}{99}
\bibitem{4-pi}
I have detected an error of double couting of the angular
factor $4\pi$ in \cite{insty}.
Thus, quoted numbers there should be divided by $4\pi$.
\bibitem{nakano}
The process of this type has been discussed with I. Nakano,
which is much appreciated.
\end{thebibliography}
\end{document}